\documentclass[10pt,a4paper]{article} 
\usepackage[margin=1.5cm]{geometry}
\usepackage[numbers]{natbib}
\usepackage{amsmath}
\usepackage{paralist}
\usepackage[hidelinks]{hyperref}
\hypersetup{colorlinks,linkcolor={blue},citecolor={blue},urlcolor={blue}}
\usepackage[utf8]{inputenc}
\usepackage{tabularx}
\usepackage{booktabs}
\usepackage{xcolor}
\usepackage{amsfonts}
\usepackage{listings}
\title{DiffSharp: An AD Library for .NET Languages%
\thanks{\textbf{Extended abstract presented at the AD 2016 Conference, Sep 2016, Oxford UK.}}}
\author{Atılım Güneş Baydin\footnote{Corresponding Author, Dept of Computer Science, National
    University of Ireland Maynooth,
    \href{mailto:gunes@cs.nuim.ie}{\texttt{gunes@cs.nuim.ie}}\newline (Current address: Dept of Engineering Science, University of Oxford, \href{mailto:gunes@robots.ox.ac.uk}{\texttt{gunes@robots.ox.ac.uk}})}
  \qquad
  \href{http://barak.pearlmutter.net}{\color{black}Barak A. Pearlmutter}\footnote{Dept of Computer Science, National University of
    Ireland Maynooth,
    \href{mailto:barak@pearlmutter.net}{\texttt{barak@pearlmutter.net}}}
  \qquad
  \href{http://engineering.purdue.edu/~qobi}{\color{black}Jeffrey Mark Siskind}\footnote{School of Electrical and Computer Engineering, Purdue
    University, \href{mailto:qobi@purdue.edu}{\texttt{qobi@purdue.edu}}}}
\date{April 2016}

\definecolor{bluekeywords}{rgb}{0.13,0.13,1}
\definecolor{greencomments}{rgb}{0,0.5,0}
\definecolor{redstrings}{rgb}{0.9,0,0}

\lstdefinelanguage{CSharp}
{
sensitive=true,
morekeywords=[1]{
abstract, as, base, break, case,
catch, checked, class, const, continue,
default, delegate, do, else, enum,
event, explicit, extern, false,
finally, fixed, for, foreach, goto, if,
implicit, in, interface, internal, is,
lock, namespace, new, null, operator,
out, override, params, private,
protected, public, readonly, ref,
return, sealed, sizeof, stackalloc,
static, struct, switch, this, throw,
true, try, typeof, unchecked, unsafe,
using, virtual, volatile, while, bool,
byte, char, decimal, double, float,
int, lock, object, sbyte, short, string,
uint, ulong, ushort, var, void},
  keywordstyle=\color{bluekeywords},
  sensitive=true,
  basicstyle=\ttfamily\small,
	breaklines=true,
  xleftmargin=\parindent,
  aboveskip=\bigskipamount,
	tabsize=4,
morecomment=[l]{//},
morecomment=[s]{/*}{*/},
morecomment=[l][keywordstyle4]{\#},
morestring=[b]",
morestring=[b]',
}

\lstdefinelanguage{FSharp}%
{morekeywords={let, new, match, with, rec, open, module, namespace, type, of, member, %
and, for, while, true, false, in, inline, do, begin, end, fun, function, return, yield, try, %
mutable, if, then, else, cloud, async, static, use, abstract, interface, inherit, finally },
  otherkeywords={ let!, return!, do!, yield!, use!, var, from, select, where, order, by },
  keywordstyle=\color{bluekeywords},
  sensitive=true,
  basicstyle=\ttfamily\small,
	breaklines=true,
  xleftmargin=\parindent,
  aboveskip=\bigskipamount,
	tabsize=4,
  morecomment=[l][\color{greencomments}]{///},
  morecomment=[l][\color{greencomments}]{//},
  morecomment=[s][\color{greencomments}]{{(*}{*)}},
  morestring=[b]",
  showstringspaces=false,
  literate={`}{\`}1,
  stringstyle=\color{redstrings},
}

\begin{document}
\maketitle
\thispagestyle{empty}
\vspace{-7mm}

\section*{Introduction}

DiffSharp\footnote{\url{http://diffsharp.github.io/DiffSharp/}} is an algorithmic differentiation (AD) library for the .NET ecosystem, which is targeted by the C\# and F\# languages, among others. The library has been designed with machine learning applications in mind \citep{Baydin2015b}, allowing very succinct implementations of models and optimization routines. DiffSharp is implemented in F\# and exposes forward and reverse AD operators as general nestable higher-order functions, usable by any .NET language. It provides high-performance linear algebra primitives---scalars, vectors, and matrices, with a generalization to tensors underway---that are fully supported by all the AD operators, and which use a BLAS/LAPACK backend via the highly optimized OpenBLAS library.\footnote{\url{http://www.openblas.net/}}
DiffSharp currently uses operator overloading, but we are developing a transformation-based version of the library using F\#'s ``code quotation'' metaprogramming facility \citep{Syme2006}. Work on a CUDA-based GPU backend is also underway.

\section*{The .NET platform and F\#}

DiffSharp contributes a much needed advanced AD library to the .NET ecosystem, which encompasses primarily the languages of C\#, F\#, and VB in addition to others with less following, such as C++/CLI, ClojureCLR, IronScheme, and IronPython.\footnote{For a full list, see: \url{https://en.wikipedia.org/wiki/List_of_CLI_languages}} In terms of popularity, C\# is the biggest among these, coming fourth (after Javascript, SQL, and Java; and before Python, C++, and C) in the 2015 Stack Overflow developer survey;\footnote{\url{http://stackoverflow.com/research/developer-survey-2015}} and again fourth (after Java, C, and C++; and before Python, PHP, and VB) in the TIOBE index for March 2016.\footnote{\url{http://www.tiobe.com/tiobe_index}}. Initially developed by Microsoft, the .NET platform has recently transformed into a fully open source endeavor overseen by the .NET Foundation, and it is currently undergoing a transition to the open source and cross platform .NET Core project,\footnote{\url{https://dotnet.github.io/}} supporting Linux, OS X, FreeBSD, and Windows.

F\# is a strongly typed functional language---also supporting imperative and object oriented paradigms---that originated as an ML dialect for .NET and maintains a degree of compatibility with OCaml \citep{Syme2006}. F\# is gaining popularity as a cross-platform functional language, particularly in the field of computational finance.
Languages that support higher-order functions are particularly appropriate for AD as the AD operators are themselves higher-order functions.
We have developed implementation strategies that allow AD to be smoothly integrated into such languages, and allow the construction of aggressively optimizing compilers.\footnote{R6RS-AD: \url{https://Functional-AutoDiff/R6RS-AD}, Stalin$\nabla$: \url{https://github.com/Functional-AutoDiff/STALINGRAD}, DVL: \url{https://github.com/Functional-AutoDiff/dysvunctional-language} \citep{Siskind2008b,Pearlmutter2008}.}
F\# allows DiffSharp to expose a natural API defined by higher-order functions, which can be freely nested and curried, accept first-class functions as arguments, and return derivative functions. The library is usable from F\# and all other .NET languages.
We provide an optional helper interface for C\# and other procedural languages.

\section*{Project organization and example code}

The code for DiffSharp is released under the GNU Lesser General Public License (LGPL)\footnote{The LGPL license allows the use of unmodified DiffSharp binaries in any (including non-GPL or proprietary) project, with attribution.} and maintained in a GitHub repository.\footnote{\url{https://github.com/DiffSharp/DiffSharp}} The user community has been engaged in the project by raising issues on GitHub and joining in the Gitter chat room.\footnote{\url{https://gitter.im/DiffSharp/DiffSharp}}

The library is structured into several namespaces. The main AD functionality is in the \texttt{DiffSharp.AD} namespace, but numerical and symbolic differentiation are also provided in \texttt{DiffSharp.Numerical} and \texttt{DiffSharp.Symbolic}. All of these implementations share the same differentiation API (Table~\ref{TableAPI}) to the extent possible, and have support both 32-bit and 64-bit floating point. (Lower precision floating point is of particular utility in deep learning.)
The \texttt{DiffSharp.Backend} namespace contains the optimized computation backends (currently OpenBLAS, with work on a CUDA backend underway). These namespaces and functionality are directly usable from C\# and other .NET languages. For making the user experience even better for non-functional languages, we provide the \texttt{DiffSharp.Interop} interface that wraps the AD and numerical differentiation functionality, automatically taking care of issues such as conversions to and from \texttt{FSharp.Core.FSharpFunc} objects.\footnote{\url{http://diffsharp.github.io/DiffSharp/csharp.html}}

Extensive documentation on the library API,\footnote{\url{http://diffsharp.github.io/DiffSharp/api-overview.html}} along with tutorials and examples, are available on the project website. The examples include machine learning applications, gradient-based optimization algorithms, clustering, Hamiltonian Markov Chain Monte Carlo, and various neural network architectures.

\section*{Key features and contributions}

\subsection*{Higher-order functional AD API}

The fundamental elements of DiffSharp's API are the \texttt{jacobianv'} and \texttt{jacobianTv''} operations, corresponding to the Jacobian-vector product (forward mode) and the Jacobian-transpose-vector product (reverse mode), respectively. The library exposes differentiation functionality through a higher-order functional API (Table~\ref{TableAPI}), where operators accept functions as arguments and return derivative functions. For instance, for a function $\mathbf{f}: \mathbb{R}^n \to \mathbb{R}^m$, the \texttt{jacobianTv''} operation, with type $(\mathbb{R}^n \to \mathbb{R}^m) \to \mathbb{R}^n \to (\mathbb{R}^m \times (\mathbb{R}^m \to \mathbb{R}^n))$, evaluates the function at a given point, and returns the function value together with another function that can be repeatedly called to compute the adjoints of the inputs using reverse mode AD.  The API also includes specialized operations (e.g., \texttt{hessianv} for Hessian-vector product) to cover common use cases and encourage modular code.
This allows succinct implementation of differentiation-based algorithms. For instance, Newton's method for optimization can be simply coded as:
\begin{lstlisting}[language=FSharp]
// eps: threshold, f: function, x: starting point
let rec argminNewton eps f x =
    let g, h = gradhessian f x
    if DV.l2norm g < eps then x else argminNewton eps f (x - DM.solveSymmetric h g)
\end{lstlisting}
Note that the caller of \texttt{argminNewton} need not be aware of what, if any, derivatives are being taken within it.

DiffSharp provides a fixed-point-iteration operator, with appropriate forward and reverse AD rules \citep{Schlenkirch2005}. The forward mode is handled by iterating until convergence of both the primal and the tangent values, while reverse mode\footnote{Currently, when using reverse mode, closed-over variables in the functional argument to the fixed-point operator should be exposed by manual closure conversion. We hope to lift this restriction soon.} uses the ``two-phases'' strategy \citep{Christianson1994}.




\begin{table}[t]
  \centering
  \scriptsize
  \renewcommand{\arraystretch}{0.9}
  \caption{The differentiation API for $\mathbb{R} \to \mathbb{R}$, $\mathbb{R}^n \to \mathbb{R}$, and $\mathbb{R}^n \to \mathbb{R}^m$ functions provided by the AD, numerical, and symbolic differentiation modules. X: exact; A: approximate; F: forward AD; R: reverse AD; F-R: reverse-on-forward AD; R-F: forward-on-reverse AD; F/R: forward AD if $n \le m$, reverse AD if $n > m$.}
  \label{TableAPI}
  \begin{tabularx}{\columnwidth}{@{}lllXlll@{}}
    \toprule
  	& Operation & Value & Type signature & AD & Numerical & Symbolic\\
    \midrule
    $f:\mathbb{R} \to \mathbb{R}$ & \texttt{diff} & $f'$ & $\color{red}{(\mathbb{R} \to \mathbb{R}) \to \mathbb{R}} \to \color{blue}{\mathbb{R}}$ & X, F & A & X\\
    &\texttt{diff'} & $(f, f')$ & $\color{red}{(\mathbb{R} \to \mathbb{R}) \to \mathbb{R}} \to \color{blue}{(\mathbb{R} \times \mathbb{R})}$ & X, F & A & X\\
    &\texttt{diff2} & $f''$ & $\color{red}{(\mathbb{R} \to \mathbb{R}) \to \mathbb{R}} \to \color{blue}{\mathbb{R}}$ & X, F & A & X\\
    &\texttt{diff2'} & $(f, f'')$ & $\color{red}{(\mathbb{R} \to \mathbb{R}) \to \mathbb{R}} \to \color{blue}{(\mathbb{R} \times \mathbb{R})}$ & X, F & A & X\\
    &\texttt{diff2''} & $(f, f', f'')$ & $\color{red}{(\mathbb{R} \to \mathbb{R}) \to \mathbb{R}} \to \color{blue}{(\mathbb{R} \times \mathbb{R} \times \mathbb{R})}$ & X, F & A & X\\
    &\texttt{diffn} & $f^{(n)}$& $\color{red}{\mathbb{N} \to (\mathbb{R} \to \mathbb{R}) \to \mathbb{R}} \to \color{blue}{\mathbb{R}}$ & X, F & & X\\
    &\texttt{diffn'} & $(f, f^{(n)})$& $\color{red}{\mathbb{N} \to (\mathbb{R} \to \mathbb{R}) \to \mathbb{R}} \to \color{blue}{(\mathbb{R} \times \mathbb{R})}$ & X, F & & X\\
    \midrule
    $f:\mathbb{R}^n \to \mathbb{R}$ & \texttt{grad} & $\nabla f$ & $\color{red}{(\mathbb{R}^n \to \mathbb{R}) \to \mathbb{R}^n} \to \color{blue}{\mathbb{R}^n}$ & X, R & A & X\\
    &\texttt{grad'} & $(f, \nabla f)$ & $\color{red}{(\mathbb{R}^n \to \mathbb{R}) \to \mathbb{R}^n} \to \color{blue}{(\mathbb{R} \times \mathbb{R}^n)}$ & X, R & A & X\\
    &\texttt{gradv} & $\nabla f \cdot \mathbf{v}$& $\color{red}{(\mathbb{R}^n \to \mathbb{R}) \to \mathbb{R}^n \to \mathbb{R}^n} \to \color{blue}{\mathbb{R}}$ & X, F & A \\
    &\texttt{gradv'} & $(f, \nabla f \cdot \mathbf{v})$ & $\color{red}{(\mathbb{R}^n \to \mathbb{R}) \to \mathbb{R}^n \to \mathbb{R}^n} \to \color{blue}{(\mathbb{R} \times \mathbb{R})}$ & X, F & A\\
    &\texttt{hessian} & $\mathbf{H}_f$& $\color{red}{(\mathbb{R}^n \to \mathbb{R}) \to \mathbb{R}^n} \to \color{blue}{\mathbb{R}^{n \times n}}$ & X, R-F & A & X\\
    &\texttt{hessian'} & $(f, \mathbf{H}_f)$ & $\color{red}{(\mathbb{R}^n \to \mathbb{R}) \to \mathbb{R}^n} \to \color{blue}{(\mathbb{R} \times \mathbb{R}^{n \times n})}$ & X, R-F & A & X\\
    &\texttt{hessianv} & $\mathbf{H}_f \mathbf{v}$ & $\color{red}{(\mathbb{R}^n \to \mathbb{R}) \to \mathbb{R}^n \to \mathbb{R}^n} \to \color{blue}{\mathbb{R}^n}$ & X, F-R & A\\
    &\texttt{hessianv'} & $(f, \mathbf{H}_f \mathbf{v})$& $\color{red}{(\mathbb{R}^n \to \mathbb{R}) \to \mathbb{R}^n \to \mathbb{R}^n} \to \color{blue}{(\mathbb{R} \times \mathbb{R}^n)}$ & X, F-R & A\\
    &\texttt{gradhessian} & $(\nabla f, \mathbf{H}_f)$ & $\color{red}{(\mathbb{R}^n \to \mathbb{R}) \to \mathbb{R}^n} \to \color{blue}{(\mathbb{R}^n \times \mathbb{R}^{n \times n})}$ & X, R-F & A & X\\
    &\texttt{gradhessian'} & $(f, \nabla f, \mathbf{H}_f)$ & $\color{red}{(\mathbb{R}^n \to \mathbb{R}) \to \mathbb{R}^n} \to \color{blue}{(\mathbb{R} \times \mathbb{R}^n \times \mathbb{R}^{n \times n})}$ & X, R-F & A & X\\
    &\texttt{gradhessianv} & $(\nabla f \cdot \mathbf{v}, \mathbf{H}_f \mathbf{v})$ & $\color{red}{(\mathbb{R}^n \to \mathbb{R}) \to \mathbb{R}^n \to \mathbb{R}^n} \to \color{blue}{(\mathbb{R} \times \mathbb{R}^n)}$ & X, F-R & A\\
    &\texttt{gradhessianv'} & $(f, \nabla f \cdot \mathbf{v}, \mathbf{H}_f \mathbf{v})$ & $\color{red}{(\mathbb{R}^n \to \mathbb{R}) \to \mathbb{R}^n \to \mathbb{R}^n} \to \color{blue}{(\mathbb{R} \times \mathbb{R} \times \mathbb{R}^n)}$ & X, F-R & A\\
    &\texttt{laplacian} & $\mathrm{tr}(\mathbf{H}_f)$ & $\color{red}{(\mathbb{R}^n \to \mathbb{R}) \to \mathbb{R}^n} \to \color{blue}{\mathbb{R}}$ & X, R-F& A & X\\
    &\texttt{laplacian'} & $(f, \mathrm{tr}(\mathbf{H}_f))$& $\color{red}{(\mathbb{R}^n \to \mathbb{R}) \to \mathbb{R}^n} \to \color{blue}{(\mathbb{R} \times \mathbb{R})}$ & X, R-F & A & X\\
    \midrule
    $\mathbf{f}:\mathbb{R}^n \to \mathbb{R}^m$ & \texttt{jacobian} & $\mathbf{J}_\mathbf{f}$ & $\color{red}{(\mathbb{R}^n \to \mathbb{R}^m) \to \mathbb{R}^n} \to \color{blue}{\mathbb{R}^{m \times n}}$ & X, F/R & A & X\\
    &\texttt{jacobian'} & $(\mathbf{f}, \mathbf{J}_\mathbf{f})$ & $\color{red}{(\mathbb{R}^n \to \mathbb{R}^m) \to \mathbb{R}^n} \to \color{blue}{(\mathbb{R}^m \times \mathbb{R}^{m \times n})}$ & X, F/R & A & X\\
    &\texttt{jacobianv} & $\mathbf{J}_\mathbf{f} \mathbf{v}$ & $\color{red}{(\mathbb{R}^n \to \mathbb{R}^m) \to \mathbb{R}^n \to \mathbb{R}^n} \to \color{blue}{\mathbb{R}^m}$ & X, F & A\\
    &\texttt{jacobianv'} & $(\mathbf{f}, \mathbf{J}_\mathbf{f} \mathbf{v})$& $\color{red}{(\mathbb{R}^n \to \mathbb{R}^m) \to \mathbb{R}^n \to \mathbb{R}^n} \to \color{blue}{(\mathbb{R}^m \times \mathbb{R}^m)}$ & X, F & A\\
    &\texttt{jacobianT} & $\mathbf{J}_{\mathbf{f}}^T$ & $\color{red}{(\mathbb{R}^n \to \mathbb{R}^m) \to \mathbb{R}^n} \to \color{blue}{\mathbb{R}^{n \times m}}$ & X, F/R & A & X\\
    &\texttt{jacobianT'} & $(\mathbf{f}, \mathbf{J}_{\mathbf{f}}^T )$ & $\color{red}{(\mathbb{R}^n \to \mathbb{R}^m) \to \mathbb{R}^n} \to \color{blue}{(\mathbb{R}^m \times \mathbb{R}^{n \times m})}$ & X, F/R & A & X\\
    &\texttt{jacobianTv} & $\mathbf{J}_{\mathbf{f}}^T \mathbf{v}$ & $\color{red}{(\mathbb{R}^n \to \mathbb{R}^m) \to \mathbb{R}^n \to \mathbb{R}^m} \to \color{blue}{\mathbb{R}^n}$ & X, R\\
    &\texttt{jacobianTv'} & $(\mathbf{f}, \mathbf{J}_{\mathbf{f}}^T \mathbf{v})$ & $\color{red}{(\mathbb{R}^n \to \mathbb{R}^m) \to \mathbb{R}^n \to \mathbb{R}^m} \to \color{blue}{(\mathbb{R}^m \times \mathbb{R}^n)}$ & X, R\\
    &\texttt{jacobianTv''} & $(\mathbf{f}, \mathbf{J}_{\mathbf{f}}^T (\cdot))$ & $\color{red}{(\mathbb{R}^n \to \mathbb{R}^m) \to \mathbb{R}^n} \to \color{blue}{(\mathbb{R}^m \times (\mathbb{R}^m \to \mathbb{R}^n))}$ & X, R\\
    &\texttt{curl} & $\nabla \times \mathbf{f}$ & $\color{red}{(\mathbb{R}^3 \to \mathbb{R}^3) \to \mathbb{R}^3} \to \color{blue}{\mathbb{R}^3}$ & X, F & A & X\\
    &\texttt{curl'} & $(\mathbf{f}, \nabla \times \mathbf{f})$ & $\color{red}{(\mathbb{R}^3 \to \mathbb{R}^3) \to \mathbb{R}^3} \to \color{blue}{(\mathbb{R}^3 \times \mathbb{R}^3)}$ & X, F & A & X\\
    &\texttt{div} & $\nabla \cdot \mathbf{f}$ & $\color{red}{(\mathbb{R}^n \to \mathbb{R}^n) \to \mathbb{R}^n} \to \color{blue}{\mathbb{R}}$ & X, F & A & X\\
    &\texttt{div'} & $(\mathbf{f}, \nabla \cdot \mathbf{f})$ & $\color{red}{(\mathbb{R}^n \to \mathbb{R}^n) \to \mathbb{R}^n} \to \color{blue}{(\mathbb{R}^n \times \mathbb{R})}$ & X, F & A & X\\
    &\texttt{curldiv} & $(\nabla \times \mathbf{f}, \nabla \cdot \mathbf{f})$ & $\color{red}{(\mathbb{R}^3 \to \mathbb{R}^3) \to \mathbb{R}^3} \to \color{blue}{(\mathbb{R}^3 \times \mathbb{R})}$ & X, F & A & X\\
    &\texttt{curldiv'} & $(\mathbf{f}, \nabla \times \mathbf{f}, \nabla \cdot \mathbf{f})$ & $\color{red}{(\mathbb{R}^3 \to \mathbb{R}^3) \to \mathbb{R}^3} \to \color{blue}{(\mathbb{R}^3 \times \mathbb{R}^3 \times \mathbb{R})}$ & X, F & A & X\\
    \bottomrule
  \end{tabularx}
\end{table}

\subsection*{Nesting}

All the AD operators can be curried or nested. For instance, making use of currying, the internal implementation of the \texttt{hessian} operator in DiffSharp is simply

\begin{lstlisting}[language=FSharp]
let inline hessian f x = jacobian (grad f) x
\end{lstlisting}
resulting in a forward-on-reverse AD evaluation of the Hessian of a function at a point.

In another example, we can implement $z = \frac{\text{d}}{\text{d}x} \left. \left( x \left( \left. \frac{\text{d}}{\text{d}y} \left( x + y \right) \; \right|_{y=1} \right) \right) \right|_{x=1}$ in F\# as

\begin{lstlisting}[language=FSharp]
let z = diff (fun x -> x * (diff (fun y -> x + y) (D 1.))) (D 1.)
\end{lstlisting}
This can be written in C\#, using \texttt{DiffSharp.Interop}, as
\begin{lstlisting}[language=CSharp]
var z = AD.Diff(x => x * AD.Diff(y => x + y, 1), 1);
\end{lstlisting}

Correctness of AD in the presence of nesting requires avoiding \emph{perturbation confusion} \citep{SiskindPearlmutter2008a}.  For instance, in the above example of nested derivatives, DiffSharp correctly returns 1 (\texttt{val z : D = D 1.0}), while an implementation suffering from perturbation confusion might return 2. We avoid perturbation confusion by tagging values to distinguish nested invocations of the AD operators. See \citep{Siskind2005,Siskind2008b,Pearlmutter2008,Manzyuk2012} for further discussion.

\subsection*{Linear algebra primitives}

One can automatically handle the derivatives of linear algebra primitives using an ``array-of-structures'' approach where arrays of AD-enabled scalars would give correct results for derivatives, albeit with poor performance and high memory consumption. This approach was used in DiffSharp until version 0.7, at which point the library was rewritten using a ``structure-of-arrays'' approach where vector and matrix types internally hold separate arrays for their primal and derivative values, and the library recognizes linear algebra operations such as matrix multiplication as intrinsic functions \citep{Giles2008}. This allows efficient vectorization of AD, where the underlying linear algebra operations can be delegated to highly optimized BLAS/LAPACK libraries.
This approach to AD with linear algebra primitives has been adopted, for example, for the GPU-based C++ reverse AD implementation of \citet{Gremse2016}.
%
%
It is interesting to note that for application domains heavily using linear algebra, such as training a neural network,\footnote{\url{http://diffsharp.github.io/DiffSharp/examples-neuralnetworks.html}} applications typically spend more than 90\% of their running time in external BLAS/LAPACK libraries. The AD library's role in a setting like this is reduced to the intelligent plumbing of primal and derivative arrays to the external library.

\section*{Benchmarks}

We provide benchmarks measuring the AD runtime overhead of the differentiation operations in the API.\footnote{\url{http://diffsharp.github.io/DiffSharp/benchmarks.html}} The code for the benchmarks is available in the GitHub repository and we also distribute a command line benchmarking tool with each release.\footnote{\url{http://github.com/DiffSharp/DiffSharp/releases}}
We intend to add memory consumption figures to these benchmarks in the upcoming release.

\section*{Current work}

\subsection*{Generalization to tensors}

DiffSharp currently provides scalar (\texttt{D}), vector (\texttt{DV}), and matrix (\texttt{DM}) types. We are working on generalizing these to an $n$-dimensional array type, with capabilities similar to those of the Torch \texttt{Tensor} class\footnote{\url{http://torch7.readthedocs.org/en/rtd/tensor/index.html}} or the NumPy \texttt{ndarray}.\footnote{\url{http://docs.scipy.org/doc/numpy-1.10.0/reference/arrays.ndarray.html}} The main motivation for this is our interest in efficiently implementing convolutional neural networks.

\subsection*{Source transformation}
The library is currently implemented using operator overloading. One of the reasons why F\# is an interesting language for AD is its advanced metaprogramming features. The ``code quotations'' feature \citep{Syme2006} allows one to programmatically read and generate abstract syntax trees of functions passed as arguments. The symbolic differentiation module in DiffSharp already makes use of code quotations. We are developing a source-transformation-based AD implementation using this feature, which should result in both speedups and simplification of the API.

\subsection*{GPU backend}

The backend interface that we defined while vectorizing DiffSharp allows us to plug in other computation backends that the user can select to run their AD code. Our current work on DiffSharp includes the implementation of a CUDA-based backend using cuBLAS for BLAS operations, custom CUDA kernels for non-BLAS operations such as element-wise function application, and cuDNN for convolution operations.

\subsection*{The Hype library}

DiffSharp will be maintained as a basis library providing an AD infrastructure to .NET languages, independent of the application domain.
In addition to setting up this infrastructure, we are interested in using generalized nested AD for implementing machine learning models and algorithms. For this purpose, we started developing the Hype library\footnote{\url{http://hypelib.github.io/Hype/}} which uses DiffSharp. Hype is in early stages of its development and is currently shared as a proof-of-concept for using generalized AD in machine learning. It showcases how the combination of nested AD and functional programming allows succinct implementations of optimization routines\footnote{\url{https://github.com/hypelib/Hype/blob/master/src/Hype/Optimize.fs}} (e.g., stochastic gradient descent, AdaGrad, RMSProp), and feedforward and recurrent neural networks. Upcoming GPU and tensor support in DiffSharp is particularly relevant in this application domain, as these are essential to modern deep learning models.

\section*{Conclusions}

Although DiffSharp started as a vehicle for conducting research at the intersection of AD and machine learning, it has grown into an industrial-strength AD solution for F\# in particular and the cross-platform .NET platform in general.
Its functional API, combined with the ability to freely nest constructs, allows for the convenient implementation of highly modular AD-using software, as seen in the Hype library.
We aim to finalize our work on the GPU backend and tensors before September 2016.
Readers are invited to refer to the online documentation and code for more in-depth information.

\section*{Acknowledgments}
This work was supported, in part, by Science Foundation Ireland grant
09/IN.1/I2637 and by NSF grant 1522954-IIS.\@
Any opinions, findings, and conclusions or recommendations expressed in this
material are those of the authors and do not necessarily reflect the views
of the sponsors.

\bibliographystyle{unsrtnat}
\begin{small}
  \setlength{\bibsep}{0.2ex}
  \bibliography{ad2016c,../sty/QobiTeX}
\end{small}

\end{document}